%% file: framework.tex
\documentclass{sig-alternate-eprint}
\global\auskip=.55pc
\usepackage[latin1]{inputenc}
\usepackage{framework}
\usepackage{graph-model}

\begin{document}

\title{An Activity-Based Model for Separation of Duty}

\numberofauthors{3}
\author{
	\alignauthor Alessandro Colantonio\\
		\affaddr{Engiweb Security}\\
		\affaddr{Roma, Italy}\\
		\email{alessandro.colantonio@eng.it}
	\alignauthor Roberto Di Pietro\\
		\affaddr{Universit\`a di Roma Tre}\\
		\affaddr{Roma, Italy}\\
		\email{dipietro@mat.uniroma3.it}
	\alignauthor Alberto Ocello\\
		\affaddr{Engiweb Security}\\
		\affaddr{Roma, Italy}\\
		\email{alberto.ocello@eng.it} }

\maketitle

\begin{abstract}
This paper offers several contributions for separation of duty
(SoD) administration in role-based access control (RBAC)
systems. We first introduce a new formal framework, based on
business perspective, where SoD constraints are analyzed
introducing the \emph{activity} concept. This notion helps
organizations define SoD constraints in terms of business
requirements and reduces management complexity in large-scale
RBAC systems. The model enables the definition of a wide
taxonomy of conflict types. In particular, object-based SoD is
introduced using the \emph{SoD domain} concept, namely the set
of data in which transaction conflicts may occur. Together with
the formalization of the above properties, in this paper we
also show the effectiveness of our proposal: we have applied
the model to a large, existing organization; results highlight
the benefits of adopting the proposed model in terms of reduced
administration cost.
\end{abstract}

\section{Introduction}

\emph{Role-based access control} (RBAC)
\cite{ansi04:standard} is a well known and recognized good
security model for enterprise access control management.
Central to the model is the \emph{role} concept (a set of
access permissions) with users assigned to roles based on
duties to fulfil. One of the main benefits related to adopting
the RBAC model is the abstraction level introduced by a role.
Roles help organizations manage complex structures and large
number of identities and permissions within their IT systems.
Indeed, a role represents an intermediate layer between
permissions (typically managed by IT staff) and users
(typically managed by business staff). This helps organizations
prevent users from accessing information at their own
discretion. In this sense, the role is a \emph{business
concept} although not the only one to be considered when
addressing access control. Other business elements, such as
business processes or organization structure, should be
included in the overall access control model. To date only a
few implementations extend the RBAC model with other business
properties.

Another important benefit related to RBAC is represented by its
simple security administration. RBAC is universally recognized
as a policy-neutral access control model in the sense that
using hierarchies and constraints, a wide range of security
policies can be expressed, including discretionary access
control (DAC), mandatory access control (MAC), and
user-specific access control
\cite{Joshi01:models,Osborn00:mac-dac}. Among all possible
aspects of security policy, \emph{separation of duty} (SoD) is
probably the most important. Alternatively indicated as
``conflict of interest'' or ``mutual exclusion'', SoD usually
refers to the identification of operations which should not be
granted to an individual user. For instance, an employee acting
as a financial manager may not be allowed to act as a financial
auditor at the same time. There are several types of frameworks
proposed in literature for SoD administration 
\cite{gligor98:formal,Nash90:sod,sandhu88:transaction,%
sandhu90:separation,kuhn97:Mutual-Exclusion,nyanchama99:graph-sod,%
ahn99:rsl,Takabi07:fuzzy,Simon97:sod}. Nevertheless, when
thousands of users, roles, and permissions have to be managed,
such frameworks do not scale well, so that administering SoD in
large-scale RBAC systems is still quite challenging. Moreover,
existing frameworks do not leverage business elements to
simplify the SoD administration in complex environments.

This paper describes a new framework for administering
separation of duty constraints, particularly suitable for
large-scale RBAC systems. The framework is based on the
following steps: first, the business processes are decomposed
into business \emph{activities}; this may be entirely done by
business staff. Then, access permissions supporting all the
activities are identified by IT staff. Potential conflicts
among activities are also provided, again by business staff,
leveraging a business perspective. Such information is used to
compute conflicts among RBAC entities---permissions, roles and
users. We show how the proposed model offers a natural way to
define SoD constraints and to reduce management complexity in
large-scale RBAC systems. The proposed model allows a wide
taxonomy of conflict types; in particular, we introduce the
\emph{SoD domain} concept as the set of data in which
transaction conflicts may occur. Introducing SoD domains makes
it possible to easily define more expressive constraints such
as \emph{object-based} separation of duty.
Finally, we have applied the proposed model on real data from a
large organization. Results confirm that our proposal greatly
reduces the administration cost required to manage RBAC
systems.

The remainder of the paper is organized as follows:
Section~\ref{sec:related-work} offers a survey of state of the
art as for SoD constraint and role administration techniques.
Section~\ref{sec:background} summarizes the main RBAC concepts
needed to formally analyze the problem. Section~\ref{sec:model}
provides a theoretical analysis of the problem, introducing
both the activity and the SoD domain concepts.
Section~\ref{sec:examples} offers a trivial usage example of
the proposed model.
Section~\ref{sec:test} shows an application of the model to a
large, actual organization. Finally,
Section~\ref{sec:conclusions} offers some final considerations.

\section{Related work}
\label{sec:related-work}

Separation of privilege is one of the founding principles for
the protection of information according to Saltzer and
Schroeder \cite{Saltzer1975:protection}. Further, Clark and
Wilson \cite{clark87:comparison} identified separation of duty
as one of the two major mechanisms that can be implemented to
ensure data integrity. At the policy level, processes can be
divided into steps, with each step being performed by a
different agent.

Several attempts to formally define separation of duty
constraints can be found in literature, especially in
role-based access control %
\cite{gligor98:formal,Nash90:sod,%
sandhu88:transaction,sandhu90:separation}. %
Simon and Zurko \cite{Simon97:sod} provide a comprehensive
classification, enumerating the different kinds of conflicts.
Kuhn \cite{kuhn97:Mutual-Exclusion} proposes mutual exclusion
of roles as a separation of duty mechanism; SoD requirements
are categorized according to the time mutual exclusion is
applied. Also Nyanchama and Osborn \cite{nyanchama99:graph-sod}
describe a way to implement various types of conflicts; they
evaluate the effect of role hierarchies in terms of their
role-graph model. Ahn and Sandhu \cite{ahn99:rsl} define the
RSL99 language for specifying separation of duty constraints.

All the aforementioned works are mainly based on the core RBAC
entities, namely permissions, roles, sessions, and users,
missing other business elements. Yet, since SoD is a business
requirement, other business aspects affecting SoD are expected
to be identified. For instance, although a user's privileges
are often granted based on the tasks the user is expected to
fulfill, the concept of tasks is usually not explicitly modeled
in access control \cite{Irwin08:task}. Perelson and Botha
\cite{Perelson2000:sod} provide a solution for the
specification of static separation of duty requirements in
role-based workflow environments. The authors identify the
impact of SoD on work process models. Then they extend the
typical RBAC model to include the notion of task. Although
tasks are taken into consideration during conflict analysis,
they are not used as elements for defining conflicts. Likewise,
Irwin et al.~\cite{Irwin08:task} introduce the task concept as
a means to determine users' privileges and use tasks to define
some security properties. Other works attempt to highlight the
importance of business information in role-based access
control, but they principally address role administration and
definition, not always considering SoD. For example, Oh and
Park \cite{Oh01:admin} highlight how, in real implementations,
users must have permissions to complete a task. Their T-RBAC
model \cite{Oh00:trbac} is an example of RBAC extension that
introduces ``task'' as a business concept. Schaad and Moffett
\cite{Schaad02:framework} propose the \emph{Alloy} language to
model organizational control principles, such as those
expressed in separation of duty, supervision and delegation.
While other important frameworks for administering RBAC
systems strive to include business elements as well %
\cite{Oh06:arbac,nyanchama99:graph-sod,Wang03:admin}, SoD
administration based on a modeling of business is not always
provided.

\section{Background}
\label{sec:background}

This section offers some of the concepts in the RBAC model
\cite{ansi04:standard} that will be used in the following. The
entities of interest for the present analysis are:
\begin{itemize}
    \item \PERMS, the set of all possible access
        permissions;
    \item \USERS, the set of all system users;
    \item $\ROLES \subseteq 2^\PERMS$, the set of all
        roles;
    \item $\UA\subseteq\USERS\times\ROLES$, the set of all
        role-user relationships;
    \item $\PA\subseteq\PERMS\times\ROLES$, the set of all
        role-permission relationships;
    \item $\RH\subseteq\ROLES\times\ROLES$, the set of
        hierarchical relationships between roles.
\end{itemize}
The symbol ``$\succeq$'' indicates an ordering operator
representing a path of direct relationships in $\RH$. If $r_1
\succeq r_2$, then $r_1$ is referred to as the child or the
\emph{senior} of~$r_2$. Similarly, $r_2$ is the parent or the
\emph{junior} of~$r_1$.

The following functions are also provided:
\begin{itemize}
\item $\assusers\colon\ROLES\to 2^\USERS$ to identify users
    assigned to a role and to none of its senior roles,
    according to \UA.
\item $\authusers\!\colon\ROLES\to 2^\USERS$ to identify
    users assigned to a role or to at least one of its
    seniors, according to \UA{} and \RH.
\item $\assperms\colon\ROLES\to 2^\PERMS$ to identify
    permissions assigned to a role and to none of its
    senior roles, according to \PA.
\item $\authperms\!\colon\ROLES\to 2^\PERMS$ to identify
    permissions assigned to a role or to at least one of
    its seniors, according to \PA{} and \RH.
\end{itemize}

In applying dynamic security policies, the \emph{session}
concept should be taken into account. Each session identifies
the set of active roles for a given user. A user may be
associated with multiple sessions at any moment in time. This
feature supports the principle of \emph{least privilege}; that
is, a user assigned to multiple roles may activate any subset
of these roles to perform his/her tasks. This may be used in
support of dynamic separation of duty policies. According to
the RBAC standard, the function $\sessroles\colon\SESSIONS\to
2^\ROLES$ identifies the set of active roles for the user
associated to the given session.

Finally, note that according to the standard, a permission is
an abstract concept that refers to the arbitrary binding of
operations and objects. This means that elements of $\PERMS$
are actually pairs $\langle o, m\rangle$, where $o\in\OBJS$
indicates the object and $m\in\OPS$ the way in which the object
is accessed---where \OBJS{} and \OPS{} indicate, respectively,
the set of objects and operations monitored by the access
control system. In this paper permissions are always considered
as a single unit, except when referred to SoD domains (see
Section~\ref{sec:domains}), where objects are used to group
permissions.

\section{Model Definition}
\label{sec:model}

This section illustrates a new SoD model, where conflicts are
not directly defined among permissions or roles, but among
\emph{business activities}. The rationale is that large
companies typically have hundreds of thousands of permissions
and roles, while possessing only a few hundred activities.
Therefore, managing activities is often easier than managing
other RBAC entities. Tasks are a natural way to think about
user actions and their contexts \cite{Irwin08:task}, making the
identification of users performing conflicting activities more
intuitive than the identification of users assigned to
conflicting roles or possessing conflicting permissions.

Most of the existing SoD models are characterized by defining
conflicts among permissions or roles. For instance, Kuhn
\cite{kuhn97:Mutual-Exclusion} proposes mutual exclusion of
roles (i.e., roles which should not be simultaneously assigned
to a user) as a separation of duty mechanism. But defining SoD
conflicts among roles could lead to inconsistent constraint
definitions. This is because user's capabilities are vested in
permissions, not roles. For instance, suppose we have roles
$r_1$, $r_2$, and $r_3$ where only $r_1$ and $r_2$ are
specified as being mutually exclusive. However, $r_2$ and $r_3$
might be assigned with the same permission set (i.e.,
$\authperms(r_2) = \authperms(r_3)$) or, more generally, they
might share the same permissions which generate conflicts with
role $r_1$ (i.e., capabilities provided by $\authperms(r_2)
\cap \authperms(r_3)$ are conflicting with capabilities
provided by $\authperms(r_1)$). According to this observation,
both pairs $\{r_1,r_2\}$ and $\{r_1,r_3\}$ would provide a user
with the same conflicting permissions. While no user could be
assigned to both $r_1$ and $r_2$ at the same time, a user could
be simultaneously assigned to $r_1$ and $r_3$ since these are
not defined as conflicting at the role level, tough providing a
user with equal conflicting permissions. This illustrates how
defining conflicts among roles can easily lead to ill-defined
SoD constraints or, even worse, allows to entirely bypass other
constraints.

According to the previous observation, the correct approach
should be to define conflicting permissions, namely sets of
permissions which should not be simultaneously possessed by the
same user. If permission conflicts are known, it will be
possible to deduce role conflicts: conflicting roles would be
those having conflicts in the union of their assigned
permissions. Therefore, specifying conflicts among permissions
provides a finer granularity as opposed to specifying conflicts
at the role level. Yet, in large organizations there could be
hundreds of thousands of permissions, leading to an
unmanageable situation. Besides this, it is usually difficult
to find a user within the organization who has both the
business and the IT knowledge required to identify conflicting
permissions.

In order to reduce complexity of SoD constraint description, we
propose an alternative approach where conflicts are not
directly defined among permissions or roles. Before explaining
how the proposed model addresses this problem, it is necessary
to introduce the \emph{activity} concept.

\subsection{Business Activities}
\label{sec:actvt-model}

Activities are identified by decomposing business processes of
an organization. From an access control point of view, an
activity is a set of permissions necessary to perform a certain
task.
In this sense, the activity and role concepts are similar in
that they both group permissions, but do have some important
differences:
\begin{itemize}
    \item \emph{Meaning}. Roles group permissions to be
        assigned to a user in order to support the work
        he/she has to do, but do not always directly map to
        specific business activities. Some roles having no
        business meaning could simply be defined out of
        convenience. For example, when the hierarchical
        RBAC model is adopted, a so called ``connector
        role'' represents the intersection of permissions
        assigned to all the derived roles; but the
        connector role may have no business meaning.
        Additionally, some RBAC implementations offer a way
        to define IT and business roles. Business roles are
        typically established by the business (e.g.,
        Employee, President, Trader, etc.)\ whereas IT
        roles are established by the application owners
        (e.g., admin, user, auditor, etc.). In these
        systems, it is expected that mapping business roles
        to IT roles would simplify policy specification. In
        such case, IT roles exist only to reduce the
        overall system administration effort, while
        business roles could allow to perform more than one
        business activity.
    \item \emph{Cardinality}. A typical large-scale
        organization could have thousands of roles defined
        in its access control system, while having no more
        than a few hundred business activities. In any
        organization, activities depend on company
        objectives, not organizational structure or
        headcount. For instance, suppose two sales
        management roles \verb|na_sales_mgr| and
        \verb|emea_sales_mgr| are assigned different
        permission sets as they are related to different
        markets. Despite the need for two distinct roles,
        they likely allow the same kind of activities to be
        performed. While new roles are created in response
        to new sales markets, no new activities will be
        required. Thus, working with activities instead of
        roles (whenever the problem permits this
        replacement) allows an organization to better
        address its growth.
    \item \emph{Abstraction}. A role is defined to manage
        permissions, while an activity is a business
        concept independent from permissions. Activity
        constraints could be identified by business staff
        who have no knowledge of access control. Instead,
        the role constraint definition could require a
        joint effort between business and~IT.
\end{itemize}

\subsection{Activity Concept Formalization}

Activities are obtained by decomposing business processes into
more elementary components, resulting in a tree structure
formally described as follows:
\begin{itemize}
    \item the set $\ACTVT$ contains all activities obtained
        by decomposing business processes;
    \item the set $\ACTVTH\subseteq\ACTVT\times\ACTVT$
        defines a partial order on the hierarchy tree;
        $\langle a_p, a_c\rangle\in\ACTVTH$ means that the
        activity $a_p$ is the parent of the activity $a_c$,
        also represented by $a_c\to a_p$;
    \item the activity tree has only one root;
    \item $\forall a\in\ACTVT$, the activity $a$ has only
        one direct parent, namely $\forall a_p,a_c\in\ACTVT
        : a_c\to a_p\implies\nexists a'_p\in\ACTVT,
        a'_p\neq a_p : a_c\to a'_p$.
\end{itemize}

Activity hierarchy is described using the concept of
\emph{generalization}, that is, using an ``is-a'' relation
\cite{moffett99:hierarchies}. Given two activities
$a_p,a_c\in\ACTVT$, then $a_c$ \mbox{``is-a''} $a_p$ indicates
that $a_p$ is more general than $a_c$. This relationship,
represented with the notation $a_c\to a_p$, also defines a
partial order; thus, $a_c\succeq a_p$ indicates the existence
of a hierarchical relationship pathway of ``$\to$'' from $a_c$
to $a_p$. Note that, without loss of generality, it is always
possible to individuate a unique root for a set of activities;
for example, a virtual activity ``collecting'' all the high
level activities of the organization could be always defined.

Each activity is supported by sets of permissions, which allow
activities to be performed.
To obtain a greater level of flexibility, the \emph{permission
grouping} concept has been introduced into the model. Instead
of assigning permissions directly to activities, permissions
are first grouped into one ore more subsets. For a user to
perform a given activity, such user must have \emph{all} the
permissions associated to at least one activity-related
grouping. This models situations where harmful activities are
not defined by a single action. If a user is assigned all the
permissions of a grouping, it is possible to assert that the
user performs the activity associated to such a grouping. More
precisely, a set of permission groupings is attached to each
activity and can be formalized as follows:
\begin{itemize}
    \item $\GRPS\subseteq 2^\PERMS$ represents the possible
        permission groupings which can be assigned to
        activities.
    \item $\ACTVTG\subseteq\ACTVT\times2^\GRPS$ expresses
        the origin of a permission grouping in a given
        activity.
\end{itemize}
The function $\actvtgrps \colon \ACTVT \to 2^\GRPS$ provides
the set of groupings associated to an activity. Given an
activity $a\in\ACTVT$, it can be formalized as
\[
    \actvtgrps(a) = \{g\in\GRPS\mid
        \exists\langle a, g\rangle\in\ACTVTG\}.
\]
The previous function can be extended to takes into account the
activity breakdown structure, namely
\begin{multline*}
    \actvtgrps^*(a) = \{g\in\GRPS\mid
        \exists a'\!\in\ACTVT: a'\!\succeq a,\\
        \langle a'\!, g\rangle\in\ACTVTG\}
\end{multline*}
provides all the groupings assigned to $a$ and its children.

Notice that activities and groupings can be seen as
specialization of roles. In particular, assume the role entity
is enriched with a new attribute making a distinction among
regular roles, activities, and groupings. Therefore,
$\ACTVT\subseteq\ROLES$ and $\GRPS\subseteq\ROLES$, ensuring
that elements in $\ACTVT$ and $\GRPS$ are unassignable. Similar
to users assigned to hierarchical-related roles, a user
performing an activity could also perform all parent activities
in the activity tree. Moreover, the set $\PA$ will be used to
assign permissions to groupings, while $\RH$ will be used to
define the activity structure with groupings as leaves of the
tree. Constraint formalization and mechanisms defined for roles
may also be used among activities and groupings.

The following sections formalize SoD conflicts within the
model. Because of the analogy between roles and activities,
such definitions can be directly derived from SSD (Static
Separation of Duty) and DSD (Dynamic Separation of Duty)
constraint definitions of \cite{ansi04:standard}.

\subsection{Conflict Definition}
\label{sec:sod-approach}

Potential SoD conflicts are identified among activities, namely
from a business perspective. The set $\SODG\subseteq 2^\ACTVT
\times \mathbb{N}$, describing activity conflicts, is a
collection of pairs $\langle A, n\rangle$ where each $A$ is an
activity set, while $n\geq 2$.
No user should perform more than $n-1$ activities in $A$
for each $\langle A, n\rangle\in\SODG$. 
Since activities are hierarchically related, users should not
perform neither activities in $A$ nor parent activities of $A$.

An observation key is that, since activities identify
permission sets, it is possible to derive conflicting
permissions from conflicting activities. For example, let us
assume the activities ``Invoice Creation'' and ``Invoice
Approval'' conflict; further, assume permissions
\{\,\verb|new_inv1|, \verb|new_inv2|, \verb|new_inv3|\,\} are
needed to complete the activity ``Invoice Creation'' and
\{\,\verb|app_inv1|, \verb|app_inv2|\,\} to complete the
activity ``Invoice Approval''. Thus, there should not be any
user possessing permissions \{\,\verb|new_inv1|,
\verb|new_inv2|, \verb|new_inv3|, \verb|app_inv1|,
\verb|app_inv2|\,\} since these permissions allow users to
perform conflicting activities, namely creating and approving
an invoice by oneself. Based on this observation, the following
are different kinds of conflicts among RBAC entities that can
be identified when adopting the proposed model.

\paragraph*{Conflicting and illegal permissions}

Permissions conflict when they allow execution of conflicting
activities. Formally, given a set $P \subseteq \PERMS$,
permissions in $P$ conflict if the following holds:
\begin{multline}
    \label{eq:conflicting-perms}
	\exists \langle A, n\rangle\in \SODG,\;
	\exists A' \!\subseteq A
	\;:\;
    |A'| \geq n
    \implies\\
	\forall a\in A'\!,\;
    \exists P'\!\subseteq P \;:\;
    P'\!\in\actvtgrps^*(a).
\end{multline}
When $|P|=1$, the permission $p\in P$ is \emph{illegal},
meaning that the permission allows the execution of conflicting
activities by itself. This happens whenever an application does
not allow the execution of different activities recurring to
different functionalities (e.g., the application offers only
one functionality for the ``Invoice Creation'' and ``Invoice
Approval'' activities).

\paragraph*{Conflicting and illegal roles}

Roles are conflicting when the union of their assigned
permissions allow execution of conflicting activities.
Formally, given a set $R \subseteq \ROLES$, roles in $R$
conflict if the following holds:
\begin{multline}
    \label{eq:conflicting-roles}
	\exists \langle A, n\rangle\in \SODG,\;
	\exists A' \!\subseteq A
	\;:\;
    |A'| \geq n
	\implies\\
	\forall a\in A'\!,\;
	\exists P'\!\subseteq\textstyle\bigcup_{r\in R}\assperms(r)
    \;:\\\quad
    P'\!\in\actvtgrps^*(a).
\end{multline}
When $|R|=1$, the role $r\in R$ is \emph{illegal}, namely it
contains conflicting or illegal permissions.
In order to consider the role hierarchy,
Equation~\ref{eq:conflicting-roles} can be easily extended
using $\authperms(r)$ instead of $\assperms(r)$. Note that
Equation~\ref{eq:conflicting-roles} is derived from
Equation~\ref{eq:conflicting-perms} by substituting the set of
all conflicting permissions $P$ with the union of permissions
assigned to roles in $R$.

\paragraph*{Conflicting and illegal users}

Users conflict when the union of permissions assigned to their
roles allows execution of conflicting activities. Formally,
given a set $U \subseteq \USERS$, users in $U$ conflict if the
following holds:
\begin{multline}
    \label{eq:conflicting-users}
	\exists \langle A, n\rangle\in \SODG,\;
	\exists A' \!\subseteq A
	\;:\;
    |A'| \geq n
	\implies\\
	\forall a\in A'\!,\;
	\exists P'\!\subseteq\textstyle
        \bigcup_{r\,\mid\,\exists u\in U\,:\, u \in\assusers(r)}\assperms(r)
    \;:\\
    P'\!\in\actvtgrps^*(a).
\end{multline}
When $|U|=1$, the user $u\in U$ is \emph{illegal}, namely it is
assigned to conflicting or illegal roles.
In order to consider the role hierarchy,
Equation~\ref{eq:conflicting-users} can be easily extended
using $\authperms(r)$ instead of $\assperms(r)$, and
$\authusers(r)$ instead of $\assusers(r)$. Note that
Equation~\ref{eq:conflicting-users} is derived from
Equation~\ref{eq:conflicting-perms} by substituting the set of
all conflicting permissions $P$ with the union of permissions
assigned to users in $U$ through their assigned roles.

Equation~\ref{eq:conflicting-users} does not take into account
sessions and can thus be used only for static SoD constraint
definition. As far as dynamic SoD is concerned, given a session
$s\in\SESSIONS$ the previous equation can be modified as
follows:
\begin{multline}
    \label{eq:conflicting-users-dynamic}
	\exists \langle A, n\rangle\in \SODG,\;
	\exists A' \!\subseteq A
	\;:\;
    |A'| \geq n
	\implies\\
	\forall a\in A'\!,\;
	\exists P'\!\subseteq\textstyle
        \bigcup_{r\in\sessroles(s)}\assperms(r)
    \;:\\
    P'\!\in\actvtgrps^*(a).
\end{multline}
According to the RBAC model, a session is related to a single
user. In order to consider role hierarchy, the previous
equation can be easily extended using $\authperms(r)$ instead
of $\assperms(r)$.

\subsection{SoD Domains and Constraint Taxonomy}
\label{sec:domains}

The introduction of the activity concept allows flexible
categorization of several kinds of SoD constraints. For
example, the constraint taxonomy proposed in \cite{Simon97:sod}
can be extended recurring to the activity concept instead of
the role concept. The first class of constraints is represented
by \emph{Operational SoD}: activities, to be considered
conflicting, must be completed in all their steps. Therefore,
the users are allowed to execute parts of, but not the entire
set of conflicting activities. From an access control
viewpoint, a step is no more than a set of permissions
\cite{Oh01:admin,Oh00:trbac}. In our model, this concept is
mapped to permission groupings. If a particular order of
execution is required, dynamic session-based mechanisms might
be implemented.

Another SoD constraint class is represented by the
\emph{Object-Based SoD} specifying that a user cannot execute
conflicting activities on the same object. This kind of
constraint is usually not directly supported by RBAC
implementations. To support it, we introduce the \emph{SoD
domain} concept in our model. For instance, a SoD domain is a
set of data on which a single user should not complete
conflicting activities. If a user can perform conflicting
activities on different data, then it is probably impossible to
configure an illegal action. A SoD domain can be a single row
on a DBMS table, or simply the set of all data accessed by a
given application. As mentioned in
Section~\ref{sec:background}, a RBAC permission is represented
by a couple $\langle o, m\rangle$, where $o$ indicates the
object and $m$ the way in which the object is accessed. This
means that given a permission, it is possible to determine in
which domains the permission operates. Consequently, SoD
domains can be formalized as follows:
\begin{itemize}
    \item $\SODD$ indicates the set of all SoD domains.
    \item $\SODDO\subseteq\SODD\times2^\OBJS$ expresses the
        origin of an object in a given SoD domain.
\end{itemize}
The function $\permdomains \colon \PERMS \to 2^\SODD$ provides
the set of SoD domains a given permission operates. This can be
computed from the sets $\SODDO$ and $\PERMS$; given
$p\in\PERMS$ the function can be defined as:
\begin{multline*}
    \permdomains(p) = \{d \in \SODD \mid
        \exists\langle d, O\rangle\in\\\SODDO,\;
        \exists m \in \OPS,\;
        \exists o \in O \;:\;
        \langle o, m\rangle = p\}.
\end{multline*}

To bring the SoD domain concept in the equations of the
previous section, it is required that all conflicting
permissions must be confined within the domains contained in set $D$.
Therefore, Equation~\ref{eq:conflicting-perms} (conflicting and
illegal permissions) becomes
\begin{multline}
	\exists \langle A, n\rangle\in \SODG,\;
	\exists A' \!\subseteq A,\;
    \exists D\subseteq (2^\SODD\setminus\{\emptyset\})
	\;:\;
    |A'| \geq n\\
    \implies
	\forall a\in A'\!,\;
    \exists P'\!\subseteq P \;:\;
    P'\!\in\actvtgrps^*(a)
    ,\;\\
    \textstyle\bigcap_{p\in P'}\permdomains(p) = D.
\end{multline}
Similarly, Equation~\ref{eq:conflicting-roles} (conflicting and
illegal roles) becomes
\begin{multline}
	\exists \langle A, n\rangle\in \SODG,\;
	\exists A' \!\subseteq A,\;
    \exists D\subseteq (2^\SODD\setminus\{\emptyset\})
	\;:\;
    |A'| \geq n\\
	\implies
	\forall a\in A'\!,
	\exists P'\!\subseteq\textstyle\bigcup_{r\in R}\assperms(r)
    \;:\\\quad\;
    P'\!\in\actvtgrps^*(a)
    ,\;
    \textstyle\bigcap_{p\in P'}\permdomains(p) = D.
\end{multline}
Equation~\ref{eq:conflicting-users} (conflicting and illegal
users without sessions) becomes
\begin{multline}
	\exists \langle A, n\rangle\in \SODG,\;
	\exists A' \!\subseteq A,\;
    \exists D\subseteq (2^\SODD\setminus\{\emptyset\})
	\;:\;
    |A'| \geq n\\
	\implies
	\forall a\in A'\!,\;
	\exists P'\!\subseteq\textstyle
        \bigcup_{r\,\mid\,\exists u\in U\,:
        \, u \in\assusers(r)}\assperms(r)
    \;:\\\quad
    P'\!\in\actvtgrps^*(a)
    ,\;
    \textstyle\bigcap_{p\in P'}\permdomains(p) = D
\end{multline}
while Equation~\ref{eq:conflicting-users-dynamic} (conflicting
and illegal users with sessions)
\begin{multline}
    \label{eq:conflicting-users-dynamic2}
	\exists \langle A, n\rangle\in \SODG,\;
	\exists A' \!\subseteq A,\;
    \exists D\subseteq (2^\SODD\setminus\{\emptyset\})
	\;:\;
    |A'| \geq n\\
	\implies
	\forall a\in A'\!,\;
	\exists P'\!\subseteq\textstyle
        \bigcup_{r\in\sessroles(s)}\assperms(r)
    \;:\\
    P'\!\in\actvtgrps^*(a)
    ,\;
    \textstyle\bigcap_{p\in P'}\permdomains(p) = D.
\end{multline}

Although defining SoD domains at the object level allows the
finest granularity, the resulting complexity could be
unmanageable, hence preventing its application. To dominate the
curse of complexity, it is often sufficient to define a domain
as the set of all data accessed by an \emph{application} or set
of applications. Most of the time, using the application to
identify data sets provides assurance that users are prevented
from executing conflicting actions.
It is important to highlight that different data does not
necessarily  mean different domains. In fact, the same data can
belong to multiple SoD domains; moreover, working on one object
can have effects over other related objects. For this reason,
it would be too restrictive to say that permissions accessing
distinct objects cannot cause conflicts between one another.
Most of the time, defining SoD domains as all data accessed by
an application allows to correctly partition conflicting
permissions.

\section{Examples}
\label{sec:examples}

\begin{figure}
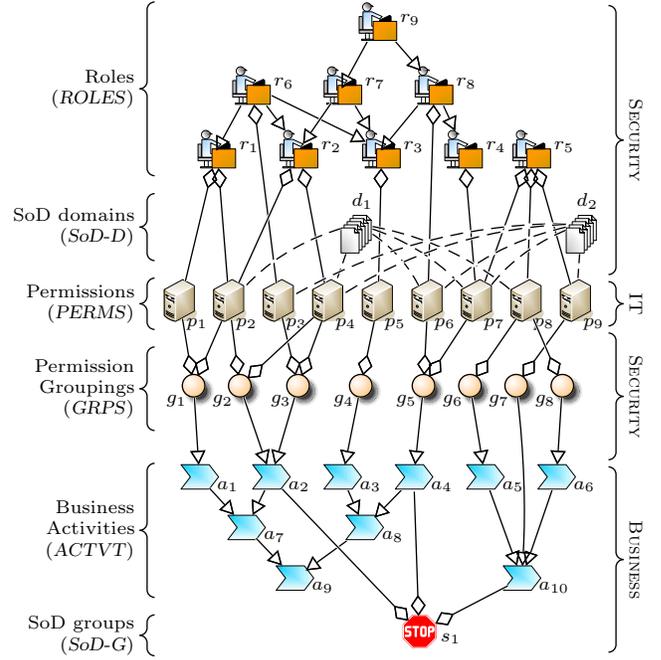

	\include{graph-model}
	\caption{Graphical representation of the SoD model}
	\label{fig:graph-model}
\end{figure}

This section introduces some examples illustrating practical
applications of the proposed SoD model. For this purpose,
Figure~\ref{fig:graph-model} shows a possible model instance,
depicting relationships among involved entities. It depicts
some example permissions ($\PERMS=\{p_1,\ldots,p_9\}$), roles
($\ROLES=\{r_1,\ldots,r_9\}$) and the corresponding
permission-role assignments. The ``aggregation'' concept is
used to assign permission sets to roles, while the
``generalization'' concept is used in defining hierarchical
relationships. For example, role~$r_1$ is a composition of
permissions~$p_1,p_2$, while role~$r_6$ is a composition of
permissions~$p_1,p_2,p_3,p_4,p_5$ since it is senior of
$r_1,r_2,r_3$ from which it inherits
permissions~$p_1,p_2,p_4,p_5$. The figure also shows how
permissions are entirely managed by IT staff, where role
definition is a typical joint-task of both business and IT
people (sometimes represented by a ``security'' staff).

Permission groupings ($\GRPS=\{g_1,\ldots,g_8\}$) are simply
aggregation of permissions, while activities
($\ACTVT=\{a_1,\ldots,a_{10}\}$) inherit permissions from
groupings or other activities. In the figure, activity~$a_2$ is
performed when a user possesses permissions~$p_2,p_4$ (i.e.,
grouping~$g_2$) or permissions $p_3,p_4$ (i.e.,
grouping~$g_3$), namely $\actvtgrps(a_2) = \{g_2,g_3\} =
\{\{p_2,p_4\}, \{p_3,p_4\}\}$. Instead, activity~$a_{10}$ is
performed when a user possesses $p_8$ (through the hierarchical
relationship with $a_5$ or $a_6$) or $p_9$ (directly through
grouping~$g_7$). In fact, permission~$p_8$ allow not only
execution of activities~$a_5,a_6$, but also activity~$a_{10}$.
Similar to a role definition task, security staff are in charge
of defining permission groupings while activity tree
identification is a typical task for business staff.

Figure~\ref{fig:graph-model} shows two distinct domains
($\SODD=\{d_1,d_2\}$). For example, permission~$p_6$ is
declared to operate in both domains, that is
$\permdomains(p_6)=\{d_1,d_2\}$, while permission~$p_2$
operates in $d_1$ but not in $d_2$. Permission~$p_1$ is not
associated to any domain, thus it can never conflict with other
permissions; in fact $\permdomains(p_1)=\emptyset$ and any
equation from Sections~\ref{sec:sod-approach}
and~\ref{sec:domains} could be satisfied. In the figure, $s_1$
is a graphical representation of a possible SoD constraint in
$\SODG$, stating that activities~$a_2,a_4,a_{10}$ cannot be
performed by the same user at the same time. According to
Section~\ref{sec:sod-approach}, this can be formally
represented by the pair~$\langle\{a_2,a_4,a_{10}\}, 3\rangle$.
Again, we emphasize the fact that defining SoD constraints is a
task for business staff.

Ignoring the SoD domain concept, $\langle\{a_2,a_4,a_{10}\},
3\rangle$ makes $r_2, r_5,r_8$ conflict because of
Equation~\ref{eq:conflicting-roles}. In fact, role~$r_2$ is
assigned with permissions~$p_2,p_4$, namely grouping~$g_2$,
thus supporting activity~$a_2$. Role~$r_5$ is assigned with
permissions~$p_7,p_8,p_9$, namely groupings~$g_6,g_7,g_8$, thus
supporting activities~$a_5,a_6,a_{10}$. Note that role~$r_5$
allows partial execution of activity~$a_4$, since permission
$p_6$ is needed to complete grouping~$g_5$. Thus, role~$r_8$ is
required to perform activity~$a_4$. Role~$r_8$ is also assigned
with permission~$p_5$ (inherited from role~$r_3$) thus allowing
execution of activity~$a_3$, that has no influence on the
analyzed SoD constraint. Permissions~$p_2,p_4,p_6,p_7,p_8,p_9$
are thus conflicting and any user possessing them (i.e.,
roles~$r_2, r_5,r_8$) would be illegal.

Introducing SoD domains, roles~$r_2, r_5, r_8$ conflict in
domain~$d_1$. In fact, permissions~$p_2,p_4,p_6,p_7,p_8$ can
operate in domain~$d_1$, while $p_9$ cannot. Although
activity~$a_{10}$ is not directly supported by $p_8$, both
activity~$a_5$ and $a_6$  satisfy
Equation~\ref{eq:conflicting-roles} since $a_5\succeq a_{10}$
and $a_6\succeq a_{10}$. Similarly, roles~$r_5, r_6, r_8$
conflict in domain~$d_2$.

Note that role~$r_9$ inherits permissions from $r_2,r_8$, thus
roles~$r_5,r_9$ conflict. For the same reason,
roles~$r_5,r_7,r_8$ conflict as well. If the previous SoD
constraint is changed to $\langle\{a_2,a_4,a_{10}\}, 2\rangle$,
role~$r_9$ will be illegal. In fact, the constraint requires
only two activities to be performed, so that $r_5$ is no longer
needed to support $a_{10}$. Adding the constraint
$\langle\{a_5,a_6\}, 2\rangle$, permission~$p_8$ will also be
illegal as well as role~$r_5$.

\section{Testing on Real Data}
\label{sec:test}

A large private company has been analyzed to highlight the
properties of the proposed SoD model. In particular, the
analyzed RBAC system is composed up of 90,287~users,
12,314~permissions (related to 67~different applications), and
16,755~roles.

In accordance to our framework, the business staff identified
an activity tree containing 298~nodes. Upon this tree,
437~conflicting activity pairs were defined. To simplify the
definition of conflicts, only conflicting activity pairs were
considered, namely constraints having the form $\langle A,
2\rangle\in\SODG$.
At the same time, 42,515~activity-permission relationships were
identified by IT professionals. To simplify this task, it was
divided among the owners of the 67~existing applications. Each
owner analyzed only permissions related to the administered
application. 
In this way each professional did not have to manage more than
a thousand activity-permission relationships. Further, one SoD
domain was defined for each application, since activities
performed by different applications adopted by the organization
do not access the same data set. Note that each identified
activity was not necessarily performed recurring to one
specific application, namely within a single domain, but it
could be performed within different contexts, thus spreading
across multiple domains.

In such a scenario, only static SoD conflicts were identified.
It was possible to deduce 4,555~illegal roles and
4,037,051~conflicting role pairs, discarding conflicts between
illegal and the remaining roles. As for roles, 2,566~illegal
permissions and 1,109,541~conflicting permission pairs were
identified, discarding conflicts between illegal and remaining
permissions. Finally, 7,047~users were found to be performing
conflicting activities, potentially able to carry out illegal
actions. Without the adoption of the proposed SoD model, such
role and permission pairs should have been directly defined,
thus requiring more effort from both the business and the IT
staff. The results of this experimental activity can be
summarized as follows: First, defining SoD conflicts through
the activity concept reduced the number of relationships to be
managed, thus leading to a reduced administration cost; in
place of defining 4,036,908~conflicting role pairs and
4,402~illegal roles, the proposed SoD model required the
definition of 437~conflicting activity pairs and
42,515~activity-permission relationships. Second, in our SoD
model the relationship identification task may be clearly
divided among business (responsible for activity conflicts) and
IT users (responsible for activity-permission relationships,
further separated among application owners). In most
organizations it is unlikely the case where users can establish
by their own all possible conflicts among roles, since this
task requires a simultaneous good understanding of both IT and
business requirements related to such roles. Further, new roles
may be defined or existing roles may be reorganized without
requiring new SoD constraints nor affecting the existing ones.

\section{Concluding Remarks}
\label{sec:conclusions}

This paper formalized a model that allows to analyze SoD
constraints adopting a business perspective. In particular,
potential SoD conflicts are defined among activities instead of
roles or permissions. In this way, it is possible to define SoD
constraints in a more natural fashion, and to reduce problem
complexity in large-scale RBAC systems.
The model also enables the definition of a wide taxonomy of
conflict types. Object-based separation of duty is introduced
using the SoD domain concept---namely the set of data in which
transaction conflicts may occur.
Experimental results supported the viability of the proposed
approach and confirmed all the claimed benefits.

To the best of our knowledge, this work represents the first
attempt to address SoD from a business perspective, yet with an
appropriate degree of formalization, that paves the way for
further research in the area.

\balance

\bibliographystyle{abbrv}
\bibliography{bibliography}

\balancecolumns

\end{document}

%% file: graph-model.tex
\scriptsize
\flushleft  
%
\SetGroupNameWidth{Organization}%
\SetRightGroupNameGap
%
\setlength{\NodeGroupDistance}{2mm}%
%
%
\LevelName{R-begin}{%
\begin{NodeGroup}
	\RoleNode{R9}{$r_9$}
\end{NodeGroup}}
\NodeGroupGap
\begin{NodeGroup}
	\hfill\hfill
	\RoleNode{R6}{$r_6$}\hfill
	\RoleNode{R7}{$r_7$}\hfill
	\RoleNode{R8}{$r_8$}\hfill
	\hfill\hfill\hfill\hfill
\end{NodeGroup}
\NodeGroupGap
\LevelName{R-end}{%
\begin{NodeGroup}
    \hfill\hfill
	\RoleNode{R1}{$r_1$}\hfill\hfill
	\RoleNode{R2}{$r_2$}\hfill\hfill
	\RoleNode{R3}{$r_3$}\hfill\hfill
	\RoleNode{R4}{$r_4$}\hfill
	\RoleNode{R5}{$r_5$}\hfill\hfill\hfill
\end{NodeGroup}}	
\NodeGroupGap
%
%
\LevelName{D-begin}{%
\LevelName{D-end}{%
\begin{NodeGroup}
    \hspace{.37\linewidth}
	\SoDDomainNode{D1}{$d_1$}
    \hspace{.37\linewidth}
    \SoDDomainNode{D2}{$d_2$}
\end{NodeGroup}}}
\NodeGroupGap
%
%
\LevelName{P-begin}{%
\LevelName{P-end}{%
\begin{NodeGroup}
	\PermissionNode{P1}{$p_1$}
	\PermissionNode{P2}{$p_2$}
	\PermissionNode{P3}{$p_3$}
	\PermissionNode{P4}{$p_4$}
	\PermissionNode{P5}{$p_5$}
	\PermissionNode{P6}{$p_6$}
	\PermissionNode{P7}{$p_7$}
	\PermissionNode{P8}{$p_8$}
	\PermissionNode{P9}{$p_9$}
\end{NodeGroup}}}
\NodeGroupGap
%
%
\LevelName{G-begin}{\begin{NodeGroup}\end{NodeGroup}}
\begin{NodeGroup}
	\SoDGroupingNode{G1}{$g_1$}
	\SoDGroupingNode{G2}{$g_2$}
    \hspace{.01\linewidth}
	\SoDGroupingNode{G3}{$g_3$}
    \hspace{.02\linewidth}
	\SoDGroupingNode{G4}{$g_4$}
    \hspace{.02\linewidth}
	\SoDGroupingNode{G5}{$g_5$}
	\SoDGroupingNode{G6}{$g_6$}
	\SoDGroupingNode{G7}{$g_7$}
	\SoDGroupingNode{G8}{$g_8$}
\end{NodeGroup}
\LevelName{G-end}{\begin{NodeGroup}\end{NodeGroup}}
\NodeGroupGap
\NodeGroupGap
%
%
\LevelName{A-begin}{%
\begin{NodeGroup}
	\ActivityNode{A1}{$a_1$}
	\ActivityNode{A2}{$a_2$}
	\ActivityNode{A3}{$a_3$}
	\ActivityNode{A4}{$a_4$}
	\ActivityNode{A5}{$a_5$}
	\ActivityNode{A6}{$a_6$}
\end{NodeGroup}}
\NodeGroupGap
\begin{NodeGroup}
    \hfill
	\ActivityNode{A7}{$a_7$}
    \hfill
	\ActivityNode{A8}{$a_8$}
	\hfill\hfill\hfill\hfill\hfill
\end{NodeGroup}
\NodeGroupGap
\LevelName{A-end}{%
\begin{NodeGroup}
	\hfill
	\ActivityNode{A9}{$a_9$}
	\hfill\hfill
	\ActivityNode{A10}{$a_{10}$}
\end{NodeGroup}}	
\NodeGroupGap
%
%
\LevelName{S-begin}{%
\LevelName{S-end}{%
\begin{NodeGroup}
	\hfill\hfill
	\SoDGroupNode{S1}{$s_1$}
    \hfill
\end{NodeGroup}}}
\begin{ArcGroup}
	\NodeBrace{R-begin}{R-end}{Roles\\ (\ROLES)}
	\NodeBrace{D-begin}{D-end}{SoD domains\\ (\SODD)}
	\NodeBrace{P-begin}{P-end}{Permissions\\ (\PERMS)} 	
	\NodeBrace{G-begin}{G-end}{Permission\\ Groupings\\ (\GRPS)}
	\NodeBrace{A-begin}{A-end}{Business\\ Activities\\ (\ACTVT)}
	\NodeBrace{S-begin}{S-end}{SoD groups\\ (\SODG)}
    \RightNodeBrace{R-begin-t}{P-begin-t}{Security}
    \RightNodeBrace{P-begin-t}{P-end-b}{IT}
    \RightNodeBrace{P-end-b}{A-begin-t}{Security}
    \RightNodeBrace{A-begin-t}{S-end-b}{Business}
	\HierArc[nodesepB=-7.5pt]{R6}{R1}
	\HierArc{R6}{R2}
	\HierArc[nodesepB=-1pt]{R6}{R3}	
	\HierArc[nodesepA=-1pt,nodesepB=-5pt]{R7}{R2}
	\HierArc{R7}{R3}
	\HierArc[nodesepB=-7.5pt]{R8}{R3}
	\HierArc{R8}{R4}
	\HierArc[nodesepB=-8pt]{R9}{R7}
	\HierArc[nodesepB=-1.5pt]{R9}{R8}
    \CompArc[nodesepA=-1.2pt]{P1}{R1}
    \CompArc[nodesepA=-1.5pt]{P2}{R1}
    \CompArc[nodesepA=-2pt]{P3}{R6}
	\CompArc[nodesepA=-2pt]{P2}{R2}
	\CompArc[nodesepA=-2pt]{P4}{R2}
	\CompArc[nodesepA=-1.4pt]{P5}{R3}
	\CompArc[nodesepA=-1pt]{P6}{R8}
	\CompArc[nodesepA=-1pt]{P7}{R4}
	\CompArc[nodesepA=-0.1pt]{P8}{R5}
	\CompArc[nodesepA=-1.5pt]{P7}{R5}
	\CompArc[nodesepA=-2.5pt]{P9}{R5}
	\ncarc[style=DashArcStyle,arcangle=-15]{D1}{P2}
	\DashArc{D1}{P4}
	\ncarc[style=DashArcStyle,arcangle=15,nodesepB=-3pt]{D1}{P6}
	\ncarc[style=DashArcStyle,arcangle=15,nodesepB=-1pt]{D1}{P7}
	\ncarc[style=DashArcStyle,arcangle=15]{D1}{P8}
	\ncarc[style=DashArcStyle,arcangle=-15]{D2}{P3}
	\ncarc[style=DashArcStyle,arcangle=-15]{D2}{P4}
	\ncarc[style=DashArcStyle,arcangle=-15]{D2}{P6}
	\ncarc[style=DashArcStyle,arcangle=-15,nodesepB=-1pt]{D2}{P7}
	\DashArc{D2}{P9}
	\HierArc[nodesepA=-1.2pt]{A1}{A7}
	\HierArc[nodesepA=-.5pt]{A2}{A7}
	\HierArc[nodesepA=-.5pt]{A3}{A8}
	\HierArc[nodesepA=-.5pt,nodesepB=-1pt]{A4}{A8}
	\HierArc[nodesepA=-1.5pt]{A7}{A9}
	\HierArc[nodesepA=-0.5pt,nodesepB=-2pt]{A8}{A9}
	\HierArc[nodesepA=-.5pt]{A6}{A10}
	\HierArc[nodesepA=-.5pt]{A5}{A10}
	\HierArc{G1}{A1}
	\HierArc{G2}{A2}
	\HierArc{G3}{A2}
	\HierArc{G4}{A3}
	\HierArc{G5}{A4}
	\HierArc{G6}{A5}
	\ncarc[style=HierArcStyle,arcangle=5]{G7}{A10}
	\HierArc{G8}{A6}
	\CompArc[nodesepA=-4pt]{P1}{G1}
	\CompArc[nodesepA=-3pt]{P2}{G1}
	\CompArc[nodesepA=-3.5pt]{P2}{G2}
	\CompArc[nodesepA=-4.5pt]{P3}{G3}
	\CompArc[nodesepA=-1pt]{P4}{G2}
	\CompArc[nodesepA=-3pt]{P4}{G3}
	\CompArc[nodesepA=-3pt]{P5}{G4}
	\CompArc[nodesepA=-3pt]{P6}{G5}
	\CompArc[nodesepA=-2.5pt]{P7}{G5}
	\CompArc[nodesepA=-2.5pt]{P8}{G6}
	\CompArc[nodesepA=-2pt]{P9}{G7}
	\CompArc[nodesepA=-4pt]{P8}{G8}
	\CompArc{A4}{S1}
	\CompArc[nodesepA=-2pt]{A10}{S1}
	\CompArc[nodesepA=-1pt,nodesepB=-2pt]{A2}{S1}
\end{ArcGroup}